# Efficient implementation of Cluster Expansion models in surface Kinetic Monte Carlo simulations with lateral interactions: Subtraction Schemes, Supersites and the Supercluster Contraction

Franziska Hess*[a,b]

a    Department of Nuclear Science and Engineering, Massachusetts Institute of Technology, 77 Massachusetts Avenue, Cambridge, MA, 02139, USA

b    Institute of Physical Chemistry, RWTH Aachen, Landoltweg 2, 52074 Aachen, Germany

* E-mail address of the corresponding author: hess@pc.rwth-aachen.de

**Abstract**

While lateral interaction models for reactions at surfaces have steadily gained popularity and grown in terms of complexity, their use in chemical kinetics has been impeded by the low performance of current KMC algorithms. The origins of the additional computational cost in KMC simulations with lateral interactions are traced back to the more elaborate Cluster Expansion Hamiltonian, the more extensive rate updating, and to the impracticality of rate-catalog-based algorithms for interacting adsorbate systems. Favoring instead site-based algorithms, we propose three ways to reduce the cost of KMC simulations: 1. Represent the lattice energy by a smaller Supercluster Hamiltonian without loss of accuracy, 2. employing Subtraction Schemes for updating key quantities in the simulation that undergo only small, local changes during a reaction event, and 3. applying efficient search algorithms from a set of established methods (Supersite Approach). The resulting algorithm is fixed-cost with respect to the number of lattice sites for practical lattice sizes and scales with the square of the range of lateral interactions. The overall added cost of including a complex lateral interaction model amounts to less than a factor 3. Practical issues in implementation due to finite numerical accuracy are discussed in detail, and further suggestions for treating long-range lateral interactions are made. We conclude that, while KMC simulations with complex lateral interaction models are challenging, these challenges can be overcome by modifying the established Variable Step Size Method by employing the Supercluster, Subtraction and Supersite algorithms (SSS-VSSM).

## 1. Introduction

The complexity of first-principles-based lattice Kinetic Monte Carlo (KMC) models to treat chemical reactions at surfaces has risen dramatically over the past few years. [1-3] While early studies were restricted to just a few elementary steps and non-interacting adsorbates on high-symmetry low-index facets, we have witnessed a significant increase in the complexity of the studied chemical reaction networks [4-13], the rise of detailed interaction models [14-21], and increased efforts to study reactions, diffusion and growth on high-index facets and stepped surfaces [7, 14, 20, 22-24]. Lateral interactions have been disregarded in kinetic models of surface reactions for a long time, in part due to the inability to obtain reliable parameters from experiments, in part due to the belief that these interactions play only a minor role in the kinetics of surface reactions, and also, in part, due to the extensive computational effort associated with computing them on the level of electronic structure theory and simulating reactions with lateral interactions in KMC. With the help of Grand Canonical Monte Carlo (GCMC) and KMC simulations, however, it has become possible for the first time to explicitly study the effect of lateral interactions and adsorbate layer ordering on the kinetics of surface reactions [17, 18, 25]. It has been thus shown that lateral interactions can have a dramatic influence on the reaction rate (several orders of magnitude), microkinetic reaction parameters (apparent activation energies [25] and reaction orders), and even the relative catalytic activity of different crystal facets [26]. Lateral interactions also give us a quantitative measure of how substitutional defects on a catalyst surface influence the reaction mechanism in the direct vicinity of the defect [1, 17] or the interaction of adsorbates with the different active sites in alloys [27, 28].

Lateral interactions in KMC simulations are often modeled in a Cluster Expansion (CE) approach [29], where the total energy of the lattice and adsorption energies are represented as a sum of interaction clusters, where each cluster represents a pattern in which adsorbates are arranged on the surface, for instance, pairwise, three-body, four body, and larger clusters. The CE is, in principle, exact, but is always truncated in practical application due to constraints in the number of parameters that can be determined (and meaningfully fitted [19]) on a first principles level. In the past years, significant effort has been spent on reducing the computational effort involved in determining the CE parameters by applying machine-learning [30, 31], and other sophisticated techniques [19, 32], and making such models more reliable [19, 31, 32].

At this point, complex Cluster Expansion models for catalytic co-adsorbate systems with several intermediates, three- and four-body interactions, and interactions "beyond nearest-neighbors" [7, 15, 17-21, 29-32] are already a reality, which makes complex CE models almost ready for practical application in the modeling of heterogeneous catalysis. However, their incorporation into KMC models is currently impeded by severe performance issues: current benchmark of two popular KMC codes report an increase of computing time per KMC step by up to four orders of magnitude after introduction of lateral interactions into a test model [2, 33]. In practical simulations, this would result in a reduction of simulated KMC time physical second by the same four orders of magnitude, making the incorporation of lateral interactions in KMC models unfeasible for reactions where we struggle to reach useful time scales (seconds to minutes) even without lateral interactions. This problem can be overcome to a certain degree by parallelization as shown by Nielsen et al. for the Zacros code [33]. However, if it was possible to improve the performance of KMC with lateral interactions, the overall computational effort would decrease, making such simulations feasible to run in serial on a workstation.

Why is KMC with lateral interactions so slow? The increased computational effort for KMC with lateral interactions essentially stems from three sources (cf. **Section 2** for details):

1.) Slow evaluation of the Cluster Expansion Hamiltonian (CEH). The CEH is essentially a sum over interaction terms, where the number of terms increases with the square of the range of lateral interactions $R$, and (in the worst case) exponentially with the order $n$ of the $n$-body interactions in the CE ($< R^{2n}$). While the energy is represented by only a single term in interaction-free models, current cluster expansion models for catalytic reactions can have 20 and more terms [7, 15, 17-19]. This means that the computational effort of every single evaluation of the Hamiltonian can be 20 times higher (or more) in CE-based models compared to models without lateral interactions.

2.) Extensive rate update due to lateral interactions. After a reaction event, the rates on all sites within the range of lateral interactions change, and must be recomputed, which requires the re-evaluation of the CEH. The number of sites within the lateral interaction range scales as $R^2$, which gives an overall scaling for 1) and 2) of $< R^{2n} \cdot R^2 = R^{4n}$.

3.) Too many different rate constants. The number of different elementary rate constants scales exponentially with the number of interaction parameters. This is a problem because the family of KMC algorithms that are the current "gold standard" in surface catalysis-related KMC [3, 13, 34-39] are efficient only for simulations in which the number of different rate constants ($M$) is lower than the number of sites ($N$). They run at fixed cost per KMC step with respect to the number of surface sites (N), which means that one can simulate arbitrarily large lattices at no additional cost (per step). Instead, their run time typically scales as $\mathcal{O}(M)$ or $\mathcal{O}(\log(M))$, and memory scales as $\mathcal{O}(M)$. However, in systems with lateral interactions, $M$ scales exponentially with the number of interaction parameters, which can be an issue even in simulations with "relatively simple" interaction models [2, 33]. In more practical models, such as our models for the HCl and CO oxidation over $RuO_2(110)$ [17], for instance, the number of distinct adsorption energies is on the order of $10^9$. This results in approximately $10^{11}$ different rate constants, which renders the rate class/type/catalog-based algorithms inadequate. Instead, site-based algorithms are called for, but these typically scale as $\mathcal{O}(N)$, i.e., they become more expensive as larger lattices are employed. Together with 1) and 2) the overall scaling becomes $< N + N + R^{4n}$ (addition, because the algorithm consists of sequential steps, see **Section 3** for details). While the site-based algorithm does not scale with the number of different elementary rate constants, the linear scaling with the number of sites is equally undesirable because large lattices must be employed to obtain meaningful results in many systems.

This raises the question whether or not it is possible to devise a KMC algorithm that works for systems with lateral interactions and runs at (nearly) fixed cost regarding both the number of elementary rate constants and the number of sites. As will be shown in the following, the overall scaling with respect to the number of sites can be brought down to $\mathcal{O}(\sqrt{N})$ in a site-based algorithm including lateral interactions, and almost fixed cost can be achieved for medium to large lattice sizes. This is accomplished through 1. a hierarchical approach to compute the Cluster Expansion Hamiltonian without loss of accuracy (Supercluster Approach), 2. realizing that the changes on the surface between two simulation steps are very small, and only the changes need to be quantified in many cases (Subtraction Schemes), and 3. employing efficient search [40, 41] and bookkeeping algorithms (Supersite Algorithm). While we do not maintain a public KMC code, we provide three sample programs (supporting material) showcasing the implementation and application of these algorithms (Supercluster, Subtraction, ) in the established VSS Method.

## 2. Kinetic Monte Carlo simulations with lateral interactions: the site-based algorithm

We employ Kinetic Monte Carlo simulations based on the Variable Step Size Method (VSSM), originally introduced as N-Fold Way by Bortz et al. [42] and as Stochastic Simulation Algorithm by Gillespie [43]. There are two basic ways to implement the VSSM, and we employ the site-based implementation, for which the formulas are given below. A comparison of the formulas relevant for the scaling for site- and rate-based implementation is given in **Table S 1** in the supporting material. Lateral interactions are modeled by a Cluster Expansion (CE) including pairwise interactions up to third-nearest neighbor (3NN) and L-shaped three-body-interaction figures. The shape of the CE employed for testing the algorithms presented here is based on our CEs for the HCl oxidation [17] and the CO oxidation [18] on RuO$_2$(110) shown schematically in **Figure 1**.

The main loop of the site-based VSSM algorithm can be quickly summarized as

1. Update: calculate rate constants for every elementary step $i$ on every site ($k_{x,y,i}$).
2. Accounting: compute the total rate $\Gamma = \sum_{x,y,i} k_{x,y,i}$.
3. Draw two random numbers $\rho_1, \rho_2$.
4. Search: select and execute the elementary step $v$, for which $\sum_{j=1}^{v-1} r_j \leq \rho_1 \Gamma \leq \sum_{j=1}^{v} r_j$.
5. Advance time by $\Delta t = -\frac{Log(\rho_2)}{\Gamma}$.

For discussing the scaling of VSSM, we can identify three distinct steps that scale with the number of sites in the simulation: 1. update, 2, accounting and 4. search. In a literal implementation (as given by the formulas above), all three of these steps scale linearly with the number of sites $N$ ($\mathcal{O}(N)$). The update step, however, is usually performed only in the immediate vicinity of a reaction event, or within the range of lateral interactions ("local update"). This means that the update step in practice scales only with the range of lateral interactions and the complexity $n$ of the cluster expansion (R$^{4n}$)

## 2.1. Cluster Expansion

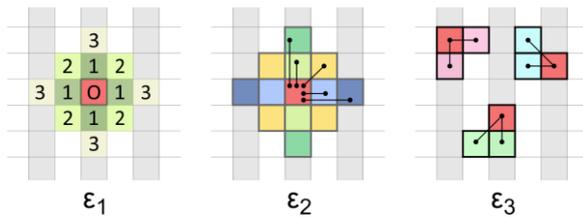

**Figure 1**: Terms of the cluster expansion as used in our models for the HCl and CO oxidation on RuO$_2$(110). The different types of sites are marked by white and grey stripes and the center site is highlighted in red. Symmetry-equivalent clusters are drawn in the same color. Left: self-interaction, or adsorption energy at zero coverage $\varepsilon_1$. The 0th, 1st, 2nd and 3rd nearest-neighbor interaction shells (0NN, 1NN, 2NN and 3NN) are indicated by numbers and colors. Center: pairwise interaction terms $\varepsilon_2$ that model the interaction between the center site (red) and one other site as considered in our model. Due to the $C_{2v}$ symmetry of the surface, the horizontal (blue) and vertical (green) interactions are inequivalent, but the diagonal (yellow) ones are equivalent by symmetry. Pairwise and self interactions amount to 12 + 1 = 13 terms. Right: The three different symmetry-inequivalent geometries for L-shaped three-body interactions. Each has three symmetry equivalents rotated by 90° (not shown), which gives 12 terms in total for the three-body interactions.

The energies of initial and final state of a reaction are calculated by a cluster expansion where the total energy $E(\sigma)$ of an adsorbate in configuration σ is represented by a sum of interaction (cluster) terms

$$E(\sigma) = \varepsilon_1(\sigma_0) + \sum_i \varepsilon_2(\sigma_0, \sigma_i) + \sum_i \sum_{j>i} \varepsilon_3(\sigma_0, \sigma_i, \sigma_j) + \cdots, \qquad \text{Eq. 1}$$

where $i$ and $j$ run over the sites in the configuration and $\varepsilon_1, \varepsilon_2, \varepsilon_3, \ldots$ denote the interaction energy for the cluster as a function of the site occupations $(\sigma_0, \sigma_i, \sigma_j, \ldots)$. While pairwise interactions can be represented by a single sum over the interaction range, three-body interactions already require a double sum. Consequently, four-body interactions will require a triple sum and so on. This is due to the number of possible interaction clusters increasing exponentially with the order of the interactions. While not all possible interaction figures are considered in practice because many terms are too small to be statistically significant, the number of terms in a CEH can be sizeable, even for short-range interactions. The CE model employed here (shown in **Figure 1**), for instance, the number of terms in Eq. 1 is 25.

## 3. Algorithms

For the overall scaling of the VSSM algorithm with CE-based lateral interactions we need to consider that the update, accounting and search steps are successive. This means that their individual scalings are not multiplied, but added to obtain the overall scaling. Using the standard site-based VSSM approach outlined in **Section 2**, the overall scaling is $\approx R^{4n}$ (Update) $+N$ (Accounting) $+N$ (Search), where $N$, $n$ and $R$ denote the number of sites on the surface, the complexity of the cluster expansion ($n$-body) and the range of lateral interactions respectively. For the overall scaling ($\mathcal{O}$), only the largest term matters, which means that the steps relevant to the overall scaling with respect to the number of sites in the lattice are the Accounting and Search steps. Both of these scales as $\mathcal{O}(N)$, and both must be improved to reduce the overall scaling. The update step, on the other hand, does not scale with $N$, but it may dominate the overall run time nonetheless.

We present several enhancements of the three main steps in VSSM, which significantly improve scaling regarding the number of sites in the lattice and the number of interaction parameters. With the algorithms outlined here, the overall scaling with respect to the number of sites becomes $\sqrt{N}$.

## 3.1. The Supercluster Approach

The Cluster Expansion Hamiltonian (CEH) in Eq. 1 is the most expensive part of a KMC calculation employing lateral interactions that can easily be responsible for 95 % and more of the total computing time. Optimization of this part is therefore the most crucial in improving the overall performance in KMC simulations with lateral interactions. To better understand the problem, we need to distinguish between cluster figures, parameters and terms in the Cluster Expansion. A cluster figure is a unique geometric arrangement of adsorbates on the surface. A variety of interaction cluster figures is shown in **Figure 1**. There can be different pairwise, three-body or higher interaction clusters, and these will usually have symmetry equivalents due to the substrate symmetry. A parameter is an energy value $\varepsilon_n$ associated with a cluster, where $n$ corresponds to the size of the interaction cluster. The parameters are the basis for the computation of the CEH. A term is a summand in Eq. 1. The number of terms is equal to the number of interaction cluster figures. The number of terms is equal to the number of addition operations as well as the number of interaction parameters that need to be retrieved from memory during the evaluation of the CEH. The number of cluster figures (and terms) is therefore the target quantity for improving the computational performance of the CEH.

The guiding principle in building a CE for an adsorbate system is in keeping the number of parameters (determined by costly first-principles methods) small, while obtaining a good fit of the adsorption energy. Since the number of possible configurations (and thereby the number of parameters) increases exponentially with cluster size, small interaction clusters (2-, 3-, and 4-body) are preferred. This approach results in a large number of terms in the cluster expansion and lengthy execution times in simulations. The guiding principle in fitting a CE is therefore in conflict with fast execution, for which the CEH should consist of as few terms as possible. To resolve this conflict, we will contract and pre-calculate parts of the sums in Eq. 1. We effectively build larger clusters ("superclusters"), thereby increasing the number of parameters, while reducing the number of terms in the cluster expansion. Of course, we do not determine the energies of these superclusters by first-principles methods, but calculate them from our 2-, 3-, and 4-body cluster expansion using Eq. 1. As will be shown, this Supercluster Approach basically trades a little bit of memory for faster evaluation and better scaling with regard to the number of interaction parameters. As this approach is in a way exactly opposite of the idea of the cluster expansion, we will call this process a "supercluster contraction".

In principle, we could pre-calculate all the energies of all the configurations σ, $E(σ)$, thereby effectively reducing Eq. 1 to a single term. However, this approach is unfeasible because the number of possible energy values is too large to store all the $E(\sigma)$ in memory (assuming that 2 GB are available). The number of energy values is equal to $n_t \cdot n_a^{N_R}$, with $N_R$, $n_t$ and $n_a$ equal to the number of sites within the interaction radius, the number site, and the number of intermediates including vacancies, respectively. For instance, in our model of the HCl oxidation over $RuO_2(100)$ with two site types (br/ot), five intermediates (including the vacancy, V/O/Cl/OH/H$_2$O), and 13 sites within the range of lateral interactions, the number of configurations distinguished by the CE is approximately $2 \cdot 5^{13} \approx 2.5 \cdot 10^9$. Storing these energy values into 32-bit floating point numbers would require $10^{10}$ bytes of memory, or 10 GB. While 10 GB is "within reach" of current technology, considering that we would like to include even longer range of interactions in our models and run them on a workstation, this solution does not appear practical.

The exact opposite of this approach would be one where no energy values are pre-calculated, and all would be computed on-the-fly, i.e., the literal evaluation of Eq. 1 for every configuration that appears during the course of the simulation. This would not require additional memory, but would be slow due to the large number of addition operations and result in bad scaling regarding the number of interaction clusters in the cluster expansion. However, there are other choices beside these limiting cases, as we can devise an approach in between that we will call Supercluster Approach.

Any fragment of the surface can be considered as a cluster in the context of the CE. These fragments can, in principle, have arbitrary size, but typical CEs contain only 1-, 2-, 3-, and 4-body interactions. Here we introduce what we call a supercluster that can be understood as any large interaction cluster that contains 3 to 6 sites. While such large interaction clusters are not contained in typical cluster expansion approaches, we realize that every supercluster contains a number of smaller 2- to 4-body interaction clusters. These can be merged into a single term that represents the interaction energy of the supercluster, which allows us to dismiss the lower interaction terms in the evaluation of the cluster expansion. While this reduces the number of addition operations required in the evaluation of the Hamiltonian, it is pure post-processing of an existing CE. It does not influence the way, or which parameters are determined in the original CE and does not incur a loss of accuracy.

The goal of the Supercluster Approach is to reduce the CE to as few terms as possible and exploiting substrate symmetry, while staying within memory limitations and keeping memory access time tractable. Beside these goals there is no general rule which superclusters should be chosen, as the choice depends on the range of interactions, the complexity of the cluster expansion and the cluster shapes considered in the CE. The only rule is that the supercluster contraction must exactly reproduce the energy of every configuration as given by Eq. 1. This means that every interaction cluster must be contained in at least one supercluster, and overlap between different superclusters must be accounted for to avoid artificial double counting of interactions. This process is exemplified in the next section.

### 3.1.1. Example with timing: 2- and 3-body interactions

As an example, we choose a CE that uses pairwise and L-shaped three-body-interactions as shown in **Figure 1** (25 terms in total). We test different choices of superclusters with timing that may give the reader an impression of what could be good choices of superclusters. As building the supercluster contraction does not require further first-principles calculations once the CE has been fitted, testing different supercluster contractions is not a computationally expensive undertaking.

We consider four different supercluster contractions, each with a different number of terms and measure their performance in computing the energy of $10^7$ configurations. It will be shown that the supercluster contraction with the smallest number of terms does not necessarily perform the best, and different speedups (compared to the literal evaluation of Eq. 1) and best performing supercluster contraction are identified for four different computing setups.

In approach A (**Figure 2**A), we treat pairwise and three-body interactions separately. For the L-shaped three-body interactions, we exploit the $C_{2v}$ substrate symmetry, and generate two superclusters with 6 sites each (blue superclusters in **Figure 2**A.I). Note that there is an area where the two superclusters geometrically overlap (white hatched sites). However, there are no L-shaped clusters that would be counted twice using this arrangement, so that this overlap is not problematic at this point. For the pairwise interactions, we can divide the interacting sites into four superclusters, where two are symmetry-equivalent (green and yellow superclusters in **Figure 2**A.II), each of which has a cluster size of 4. In addition, we need to consider the adsorption energy at zero coverage, which is a simple one-site (super)cluster. The Supercluster Contraction A (SCA) thereby reduces the CE from 25 to seven terms.

This is already a great improvement compared to the literal evaluation of the CE, backed by the performance results shown in **Section 3.1.2**, but we can tweak it some more to reduce the number of terms even further without increasing the supercluster size (SCB). First of all, we realize that the center site (red cluster in **Figure 2**A) is contained in every supercluster, so that adsorption energy at zero coverage does not need to be considered as a separate supercluster. We can either add it to a specific supercluster (weighted by ½ because the symmetry-equivalent superclusters in $C_{2v}$ symmetry always come in pairs) or add a fraction of it to every supercluster. Both options are equivalent in terms of computational effort because they only change the numbers involved, but not the supercluster figures. Similarly, we realize that the superclusters that are used exclusively for the L-shaped three-body interactions in Supercluster Contraction A contain some pairwise interactions where the superclusters A.II and A.I overlap. These sites can be included in the supercluster I without changing its shape, as displayed in **Figure 2**B.I. That would allow us to remove some of the nearest and second-nearest neighbor sites from shape II. However, we must remember that the three-body clusters actually overlap at two sites (shown by white hatching). Including pairwise interactions here would lead to double-counting of these interactions, which means we must either not include the pairwise interactions in the overlapping area or weigh them by ½. For SCB we choose the first option, while the second will be explored in SCC. Shape II has now shrunk considerably to three sites each. This means we can actually combine pairs of these superclusters again into two larger symmetry-equivalent superclusters as shown in **Figure 2**B.II (green cluster), each of which contains five sites. With this, SCB contains only four terms, which is again a significant improvement over SCA (seven terms).

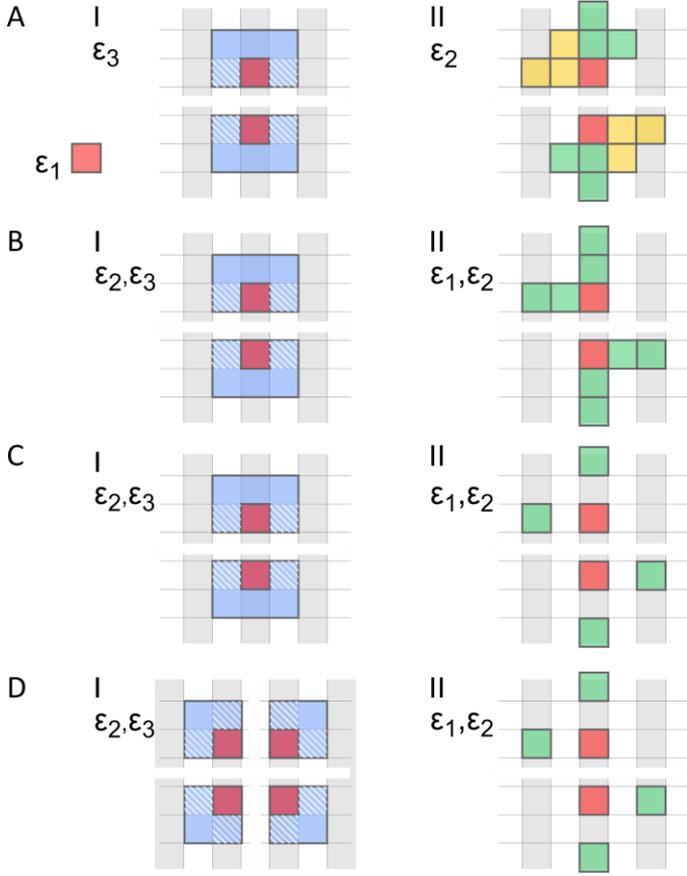

**Figure 2**: Supercluster contractions SC A, B, C, D. The center site is marked in red and symmetry-inequivalent superclusters are shown in different colors. Overlapping sites between the superclusters are hatched in white.

But this is still not the limit. Similarly to how we have included the zero-coverage adsorption energy into all the clusters, we can build SCC by adding the pairwise interaction energies where the superclusters from **Figure 2**B.I overlap into these superclusters by weighting them by ½. The cluster shape again stays the same as shown in **Figure 2**C.I, but it contains even more terms than before. Again we can remove these sites from the superclusters in **Figure 2**B.II and obtain smaller three-site superclusters as shown in **Figure 2**C.II. Now the individual sites in the cluster shape II are not connected anymore, but although most CEHs do employ mostly connected sites for the higher-order terms, this is not a requirement in the CE, and neither is it required by the Supercluster Approach. SCC consists of only 4 terms, and we can write the energy as

$$E_{SCC}(\sigma) = E_{C.I\ up}(\sigma) + E_{C.I\ down}(\sigma) + E_{C.II\ up}(\sigma) + E_{C.II\ down}(\sigma), \quad \text{Eq. 2}$$

where $E_{C.I\ up}(\sigma)$, $E_{C.I\ down}(\sigma)$, $E_{C.II\ up}(\sigma)$ and $E_{C.II\ down}(\sigma)$ stand for the two symmetry equivalents (up, down) of the C.I and C.II superclusters in **Figure 2**C.

However, larger superclusters are not necessarily better for computational speed because memory requirements and access times increase with dimension and size of the arrays. We therefore propose a fourth model where we break down the large C.I superclusters into four symmetry-equivalent smaller ones shown in **Figure 2**D.I, while keeping the second supercluster from the C model the same (**Figure 2**D.II). This increases the total number of terms to be evaluated to 6, and since we are creating additional overlap between the four D.I superclusters, we need to weigh the pairwise interaction in the overlap zone by ½ as we already did in model C. We can write the Hamiltonian for SCD as

$$E_{SCD}(\sigma) = E_{D.I.1}(\sigma) + E_{D.I.2}(\sigma) + E_{D.I.3}(\sigma) + E_{D.I.4}(\sigma) + E_{D.II\ up}(\sigma) + E_{D.II\ down}(\sigma),$$ 

Eq. 3

where $E_{D.I.1}(\sigma)$, $E_{D.I.2}(\sigma)$, $E_{D.I.3}(\sigma)$, $E_{D.I.4}(\sigma)$, $E_{D.II\ up}(\sigma)$ and $E_{D.II\ down}(\sigma)$ stand for the four symmetry equivalents (c.f. **Figure 2**D.I) of the D.I and the two symmetry equivalents of the D.II superclusters in **Figure 2**D.

### 3.1.2. Performance

The computational performance of the literal cluster expansion and supercluster models was measured by timing the energy computation of $10^7$ random configurations on four different computing setups. The results of the performance test for the three supercluster contractions presented in **Section 3.1.1** compared with the literal implementation (Eq. 1) are shown in **Table 1**. The overall time spent is in the range of seconds on all tested setups, which means that we can calculate approximately $10^7$ energies per second using the supercluster approach.

The Supercluster Approach requires some additional memory for storing the precalculated supercluster energies. We have calculated the additional memory requirement by considering the number of sites ($n_t = 2$ on RuO$_2$(110)), the number of adsorbates including vacancies ($n_a = 5$) and the supercluster sizes $s_i$, assuming that the energies are stored as single-precision floating point (32 bit). With $i$ an index that runs over all superclusters, the required memory is computed as

$$\frac{mem}{byte} = 4 \cdot n_t \cdot \sum_i n_a^{s_i}.$$

Eq. 4

Supercluster Contraction A (SCA) contains one cluster with six sites, two clusters with five sites and one cluster with one site, which requires 175 kB of memory (**Table 1**). The sizes of SCB and SCC are calculated in the same fashion, yielding 130 kB and 126 kB, respectively. SCD requires the least amount of memory because of the small size of the superclusters and efficient use of symmetry (6 kB). While we do observe some memory conservation when going from SC A to SC B/C and SC D, the required memory is less than 1 MB in all cases, which makes the additionally consumed memory negligible for small supercluster sizes.

Table 1: Performance overview of the three Supercluster Contraction models from Section 3.1.1, compared with the literal implementation. The relative speedup as given by Eq. 5 for the SC models compared to the literal evaluation of the CEH (Eq. 1) was measured on different computing setups for the computation of the energy of $10^7$ configurations. The largest speedup for each setup is highlighted in bold.

| | terms | memory / kB | relative speedup | | | |
|---|---|---|---|---|---|---|
| machine | | | Yildiz group - Fractal | | NERSC - Cori | |
| architecture | | | Nehalem EP | | Haswell | KNL |
| compiler | | | ifort 11.1 | gfortran 7.1.0 | ifort 18.0.1 | ifort 18.0.1 |
| literal | 25 | -- | 1 | 1 | 1 | 1 |
| SCA | 7 | 175 | 3.01 | 4.48 | 1.86 | 6.21 |
| SCB | 4 | 130 | 3.71 | 6.35 | 2.32 | 11.32 |
| SCC | 4 | 126 | **4.72** | **7.53** | 2.67 | **13.51** |
| SCD | 6 | 6 | 4.52 | 6.20 | **6.83** | 13.05 |
| best total time | | | 0.82 s | 2.34 s | 0.34 s | 2.81 s |

Pre-calculation of the supercluster energies during program initialization requires the literal evaluation of the cluster expansion for all possible configurations of the superclusters. In theory this time should be proportional to the required memory; however, we found the pre-calculation time too short to be even measurable reliably (< 1 ms) in our test, which means that the pre-calculation time is negligible compared to the total run time of a meaningful KMC simulation.

The relative speedup for the different supercluster contractions was determined using a toy program (`supercluster.f90`, supplied in the supporting material of this paper) compiled with ifort version 11.1 and gfortran version 7.1.0 on the Yildiz group's Fractal cluster (Intel Xeon E5530 "Nehalem-EP") and using ifort 18.0.1 on two different architectures of NERSC's Cori cluster (Intel Xeon E5-2698 v3 "Haswell" and Intel Xeon Phi 7250 "Knights Landing") (**Table 1**). While the relative performance of different compilers and CPU architectures is not within our primary area of interest, we use these four computing setups to showcase that the optimal choice of supercluster contraction may depend on the computational setup (CPU architecture, operating system, compiler etc.). In order to make the speedup of the energy evaluation more comparable across different systems, we determine a relative speedup

$$\text{rel. speedup} = \frac{t_{SC} - t_0}{t_{lit} - t_0}, \qquad \text{Eq. 5}$$

Where $t$, $t_{lit}$ and $t_0$ represent the execution time for the supercluster contraction, the execution time for the literal CE (Eq. 1) and a reference time. Subtraction of the blind time is necessary to make the true speedup of the energy calculation apparent, as our tests have shown that retrieving the random configurations from memory can be more time-consuming than performing the actual energy evaluation, and this time varies drastically between different setups. The blind time is measured by employing an energy function that returns 0 without performing any actual computation (function `energy_blind` in the sample program). The times $t, t_{lit}$ and $t_0$ were measured for computing the energy of $10^7$ random configurations.

For all tested compilers and CPU architectures, all SCs show a significant speedup (between 3.01 and 13.51) compared to the literal implementation. The trend observed for the literal evaluation and models SCA and SCB is according to the expectation that reducing the number of terms in the Supercluster Hamiltonian speeds up the computation. Since the number of terms is the same in SCB and SCC, the observed further performance boost can be explained only by the fact that the supercluster C.II (3 sites) is smaller than B.II (5 sites), which reduces memory access time. For the same reason, SCD is almost as fast as SCC on Fractal (both ifort and gfortran) and KNL, despite the fact that SCD has twice as many terms as SCC. But more surprisingly, SCD outperforms SCC on the Haswell architecture on NERSC's Cori cluster. The main difference between SCC and SCD is the size of the cluster that is used to represent the L-shaped interactions (6 in SCC and 4 in SCD). As indicated by the theoretical added memory requirement in **Table 1**, the size reduction reduces the number of different superclusters that need to be stored and retrieved. Similar to the trend observed between SCC and SCD, the performance boost is due to reduced memory access time, which appears to overcompensate the increased number of addition operations on some architectures. This suggests that with increasing supercluster size, memory access time, rather than number of addition operations may limit the actual speed of computation for some setups. The overall best performance is achieved on a Cori Haswell node with ifort 18.0.1, where the whole computation of $10^7$ energies (zero time not subtracted) was completed within 0.34 s, while the same took eight times as long (2.34 s) on KNL. However, the largest speedup is also observed for the KNL architecture, which suggests that the performance on KNL is extremely sensitive to the way the CE is computed and may require a dedicated solution to reach its full potential.

### 3.1.3. Scaling

There are many possible Supercluster Contractions for any given Cluster Expansion, and there may be no general rule for which SC performs the best. Our results indicate that computing times may be limited by memory access time when larger superclusters are employed, which suggests that the available amount of memory on current computing architectures does not pose a limitation. Rather, the exemplary results given here showcase that finding a good compromise between supercluster size and number of terms in the energy evaluation is mandatory in order to achieve the best possible performance, and several different options should be tested for the given computational setup and CE before entering the production stage in a KMC project. If different computing setups are available, it is advised to test different options as the overall performance can vary drastically even for a serial code. The reader is invited to use and modify the test program provided in the supporting material for this purpose.

Since there is an optimum cluster size, the SC is expected to scale with the range of lateral interactions ($R^2$), but not necessarily with the complexity $n$ of the CE as it is the case with the literal evaluation of the CE ($R^{2n}$). This $R^2$ scaling is rather insignificant for short-range interactions, however, and we expect a significant increase in computational effort only for longer-range interactions or surfaces with high site densities. For treatment of long-range interactions, we propose a modified approach for energy calculations presented in **Section S1.2** in the supporting material.

### 3.2. The Subtraction Scheme

Another critical step in the optimization of KMC simulations is the accounting step, where the sum of rate constants in computed:

$$\Gamma = \sum_{x,y,i} r_{x,y,i}. \qquad \text{Eq. 6}$$

In serial KMC, only one reaction event can happen at a time. Without lateral interactions, we realize that the overall change of the configuration is quite small and very localized. And even with (short-range) lateral interactions, the interacting zone is usually negligible compared to the whole lattice. This fact is exploited local update scheme (**Figure 3a**), which means that reaction rates are recomputed only in the vicinity of a reaction event or within the radius of lateral interactions. This approach is used in all current KMC codes to our knowledge. Extending the idea of the local update scheme to other important quantities, such as Γ, allows us to develop new algorithms that utilize the fact that the change between subsequent configurations is small. Another algorithm for efficient recalculation of adsorption energies with long-range lateral interactions in sketched in **Section S1.2** in the supporting material.

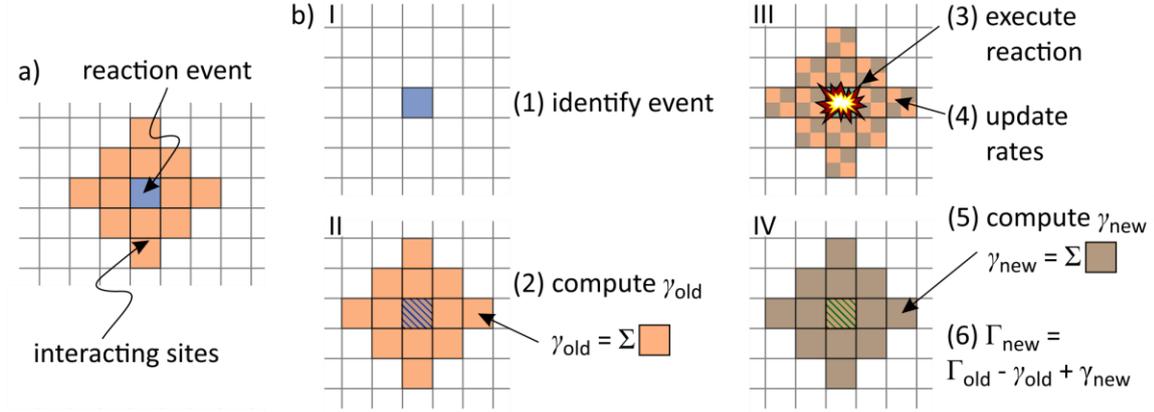

**Figure 3**: a) Local update scheme. Although the whole lattice is much bigger, only the interacting sites (orange) around the reaction event (blue) need to be updated. b) Algorithm of the subtractive Γ update scheme.

In the first simulation step, Γ is computed by Eq. 6, which is $\mathcal{O}(N)$ regarding the number of lattice sites. After a reaction event, only the rates within the range of lateral interactions change, as indicated by the orange zone in **Figure 3a**. The recalculation of Γ then becomes an $\mathcal{O}(1)$ algorithm if we compute only the change of Γ between the two steps, rather than recomputing it from scratch, as outlined below.

Denoting the site of the reaction event identified in the search step (blue site in **Figure 3b.I**) as $(x_0, y_0)$, the range of lateral interactions as $R$, and $\gamma_{old}$ and $\gamma_{new}$ the local sum of reaction rates around the reaction event (orange/brown zone in **Figure 3b.II/IV**) before and after the change, the (local) $\gamma_{old}$ is computed by taking the sum of rates around the reaction event (including the event itself) *before* the event happens (**Figure 3b.II**):

$$\gamma_{old} = \sum_{(x-x_0)^2+(y-y_0)^2 \leq R^2} \sum_i r_{x,y,i}. \qquad \text{Eq. 7}$$

Then, after executing the reaction event (**Figure 3b.III**) and locally updating the rates ($r'_{x,y,i}$), we compute $\gamma_{new}$, again only in the zone around the reaction event (brown area in **Figure 3b.IV**) using the updated rates:

$$\gamma_{new} = \sum_{(x-x_0)^2+(y-y_0)^2 \leq R^2} \sum_i r'_{x,y,i}. \qquad \text{Eq. 8}$$

Finally, Γ is updated:

$$\Gamma_{new} \rightarrow \Gamma_{old} - \gamma_{old} + \gamma_{new} \qquad \text{Eq. 9}$$

As the calculation of $\gamma_{old}$ and $\gamma_{new}$ do not contain a sum over all sites, but rather only over the sites within the interacting range, this accounting algorithm does not scale with the total number of sites in the lattice, but only with the interaction range ($R^2$). The Subtraction Scheme can face numerical stability issues in systems where the highest and lowest rate constants differ by many (> 10) orders of magnitude (stiff systems). While this usually does not have noticeable consequences for the results of KMC simulations, it requires an error detection and handling strategy that is presented in **Section S1.1** in the supporting material.

### 3.2.1. Performance

A toy program to test this algorithm is provided in the supporting material with instructions (`subtraction.f90`). The total runtime using the total and subtractive $\Gamma$ computation Scheme without performing search for lattice sizes up to ($987 \times 987$) over $10^5$ steps measured on Fractal (ifort 11.1) is shown in **Figure 4** in a log-log plot.

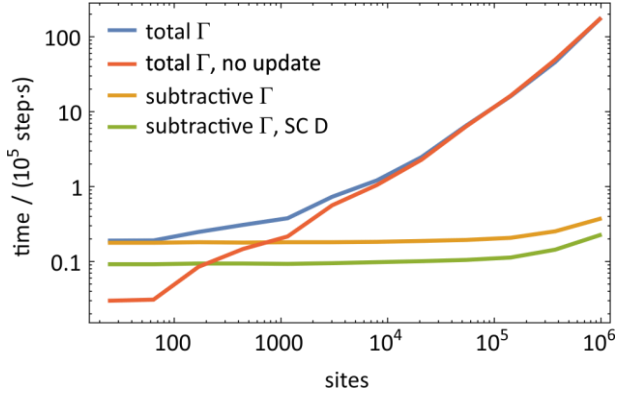

**Figure 4**: Timing of the Subtraction Scheme to evaluate $\Gamma$ (without search), using local update. The CE was evaluated literally (Eq. 1) in all cases except green, where SCD was used.

The run time of the local update scheme with total $\Gamma$ evaluation (shown in blue in **Figure 4**) increases linearly with the number of sites, although the expected $\mathcal{O}(N)$ scaling is observed only for larger lattices (> $10^4$ sites). For smaller sizes, the run time is governed the update step, which is slower than the accounting step, but has a constant run time. By deactivating the rate update (keeping only the evaluation of $\Gamma$ by Eq. 6 active), we can make the true scaling of the evaluation of $\Gamma$ apparent, which is $\mathcal{O}(N)$ as indicated by the slope of 0.98 of the red line in **Figure 4**.

The run time of the local Subtraction accounting where $\Gamma$ is evaluated according to Eq. 9 is shown in orange in **Figure 4**, and we observe that the run time of the local subtraction scheme remains constant up to $10^5$ sites. The slight increase for larger lattice sizes originates from the lattice initialization, which scales as $\mathcal{O}(N)$ and was not subtracted for the purpose of this test.

Of course, the local subtractive accounting scheme can be combined with the Supercluster Approach presented in **Section 3.1** to get even better run times. Without re-examining all possible SCs, we provide the timing only for SCD, combined with the local subtractive accounting scheme in the green curve in **Figure 4**. Since the SC significantly speeds up the update step, we observe that the runtime decreases for all lattice sizes, while scaling does not change. While this decrease looks modest in the log-log plot, the Supercluster Contraction reduces the total run time by 40-50 % in this test. This is less than the speedup of the energy function alone reported in **Section 3.1.2**, which is simply due to the fact that there are several successive steps in the algorithm with constant run time, and the energy calculation is only a part of that. Furthermore, our test performs only one energy computation per step and updated site, which is comparable to unimolecular desorption in surface catalysis. A "real" KMC simulation, however, will involve bimolecular surface reactions in addition to adsorption and desorption, which requires additional energy computations for initial and final states on two sites. It is therefore safe to assume that our test underestimates the total speedup due to the Supercluster Contraction compared to a real KMC simulation.

### 3.3. The Supersite Search

In the site-based implementation of the VSSM algorithm, the search step scales with the number of sites in the simulation. This step is quite cheap compared to the update and accounting step, so that performance improvements here have a comparably small effect on the total computational time (cf. **Section 3.4**). However, while the accounting and update steps can be reduced to $\mathcal{O}(1)$ with the previously presented new algorithms, the scaling of the search step cannot be improved beyond $\mathcal{O}(\log(N))$, which is achieved by a binary search algorithm. Although binary search offers the best possible scaling, we have found that maintaining and updating a tree structure throughout all the steps of the VSSM algorithm is quite slow compared to less sophisticated algorithms. As will be shown in **Section 3.4**, however, the run time is governed by the update step in simulations with lateral interactions even for medium to large lattice sizes, so that we consider the binary search an attractive option only for very large simulations.

Instead, we favor the 2-Level Method introduced by Maksym for the KMC simulation of MBE growth [41], which offers both good scaling and speed. It scales approximately as $\mathcal{O}(\sqrt{N})$, but can be further refined into the K-Level method ($\mathcal{O}(N^{1/K})$) proposed by Blue et al. [40]. We propose minor alterations to the 2-Level Method in order to combine them efficiently with the previous algorithms, so we will review the key ideas of the 2-Level (or Supersite) algorithm. A scaling test and sample code (`supersite.f90`) for our model with lateral interactions for large simulation lattices will be provided. We will show that this traditional algorithm can be efficiently combined with a Subtraction Scheme in the accounting step to give even better overall performance than the original 2-Level Method.

Maksym's 2-Level Algorithm involves geometrically dividing the surface into chunks [41] that we will call "supersites", each of which contains a predefined number of lattice sites, and storing the sum of rates, $\Gamma$, separately for each supersite, which allows for faster search and accounting, both of which ideally scale as $\mathcal{O}(\sqrt{N})$ in the original scheme. By combining it with the Subtraction scheme described in **Section 3.2**, the scaling of the accounting step can be further improved to $\mathcal{O}(1)$.

### 3.3.1. Approach

The surface consists of ($n_x \cdot n_y$) sites. In the standard site-based approach (**Figure 5**a), each site (*x, y*) is considered individually in the computation of $\Gamma$. In the search step (linear search), sites are visited successively (indicated in light blue in **Figure 5**a) and the coordinates (*x,y,i*) are reduced to a single index *j*, with each *j* corresponding to an individual elementary step. For each elementary step *j*, the condition

$$\sum_{j=1}^{\lambda-1} r_j \leq \rho_1 \Gamma \leq \sum_{j=1}^{\lambda} r_j \qquad \text{Eq. 10}$$

is assessed until the reaction event ($\lambda$, marked by ×) is identified (32 comparisons in the example). In the supersite algorithm, we divide the surface into supersites or chunks, each of which contains a number of sites as shown in **Figure 5**b.

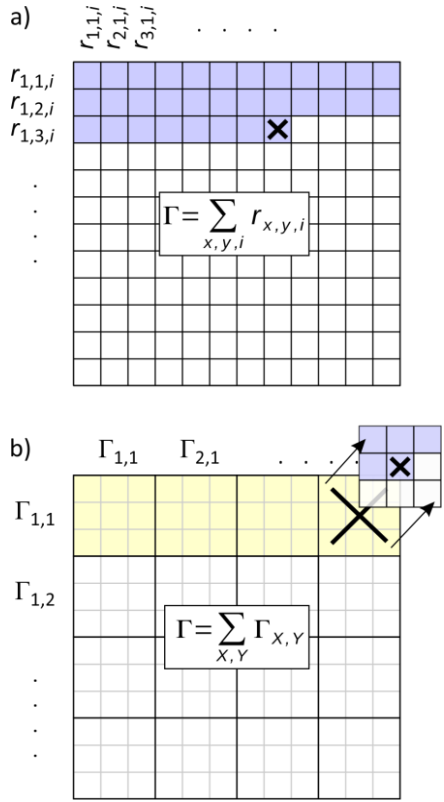

**Figure 5**: Scheme of the surface division and search in the supersite approach. a) Linear search. Sites visited are marked in blue, reaction event is marked by × (32 comparisons). b) Supersite search. Examined supersites are marked in light yellow, identified supersite is marked by ×. Then linear, search is performed within the supersite (indicated in light blue, 9 comparisons).

Each supersite $(X,Y)$ stores the sum of rates within the supersite ($\Gamma_{X,Y}$), where $s_x$ and $s_y$ denote the size of the supersite and $i$ the $i$th elementary step on site $(x,y)$:

$$\Gamma_{X,Y} = \sum_i \sum_{\substack{x= \\ (X-1)s_x}}^{X \cdot s_x} \sum_{\substack{y= \\ (Y-1)s_y}}^{Y \cdot s_y} r_{x,y,i}. \qquad \text{Eq. 11}$$

This approach allows us to store parts of Γ separately in successive steps. This enables a more efficient search algorithm (**Section 0**). While the update step can be performed as usual, Γ is calculated in the accounting step by summing over supersites, either via the standard algorithm (total Γ, $\mathcal{O}(\sqrt{N})$):

$$\Gamma = \sum_{X,Y} \Gamma_{X,Y}. \qquad \text{Eq. 12}$$

Or, again a subtraction scheme similar to Eq. 9 ($\mathcal{O}(1)$) can be employed by first subtracting the $\gamma_{old,i}$ from the previous step ($i$ runs over the $(X,Y)$ within the range of lateral interactions) that include interacting sites from $\Gamma_{old}$ from the previous step, and then adding the updated $\gamma_{new,i}$ to obtain the updated Γ:

$$\Gamma \to \Gamma_{\text{old}} - \sum_i \gamma_{old,i} + \sum_i \gamma_{new,i}. \qquad \text{Eq. 13}$$

### 3.3.2. Search

As in the standard VSSM algorithm, the search step starts with the selection of a random number $\rho_1$. Employing the rate sums of the super sites $\Gamma_{X,Y}$, the search step is broken into two parts. First, we examine the supersites (indicated in yellow in **Figure 5**b) with coordinates $(X, Y)$, reduced to a single index, $J$. The supersite $\Lambda$, that contains the reaction event (marked by the large × in **Figure 5**b) is identified by the condition

$$\sum_{J=1}^{\Lambda-1} \Gamma_J \leq \rho_1 \Gamma \leq \sum_{J=1}^{\Lambda} \Gamma_J. \qquad \text{Eq. 14}$$

Then we refine the search by examining only the sites located within the supersite $\Lambda = (X, Y)$ (highlighted in blue in **Figure 5**b). Now the index $j$ runs only over the $(x,y,i)$ values located within the supersite, i.e., $(X-1)s_x + 1 \leq x \leq X \cdot s_x$, and $(Y-1)s_y + 1 \leq y \leq Y \cdot s_y$, so that

$$\Gamma_{J-1} + \sum_{j=1}^{\lambda-1} r_j \leq \rho_1 \Gamma \leq \Gamma_{J-1} + \sum_{j=1}^{\lambda} r_j. \qquad \text{Eq. 15}$$

The site $\lambda$ thus identified is marked by a small × in **Figure 5**b. In the example in **Figure 5**, the linear search requires 32 comparisons, while the supersite search requires only 9.

The reader may notice that the two search algorithms point to different sites in **Figure 5**a and b. This is due to the fact that different search algorithms visit the sites in a different sequence. While the linear search usually operates column-by-column or row-by-row, the supersite search divides the surface into square (or rectangular) chunks visited successively. The two algorithms thus create a different event chain for the same random seed, but the simulation results will (on average) be the same, no matter the sequence the sites are visited. The performance of the supersite algorithm is sensitive to the choice of the chunk size $s_x$ and $s_y$, and the optimal $\mathcal{O}(\sqrt{N})$ performance is observed only for $s_x \approx \sqrt{n_x}$ and $s_y \approx \sqrt{n_y}$ (see **Section S2** in the supporting material and Ref. [41] for details).

### 3.4. Overall performance

All the presented algorithms can, and should be combined to obtain the best possible performance as shown in the final scaling test in **Figure 6**. We differentiate five cases combining different algorithms. First, the case where we evaluate Γ by Eq. 6 (total Γ) and perform a linear search (blue curve in **Figure 6**). Since both the accounting and search step scale as $\mathcal{O}(N)$, the overall run time also follows this scaling. This remains true if we replace either the search step by the Supersite Search (orange curve in **Figure 6**) or the accounting step by the Subtraction Scheme (green curve in **Figure 6**). Although both methods improve overall performance, the $\mathcal{O}(N)$ scaling remains the same. The Subtraction Scheme for the evaluation of Γ offers a greater boost in performance, and is always worth implementing, even in KMC simulations without lateral interactions.

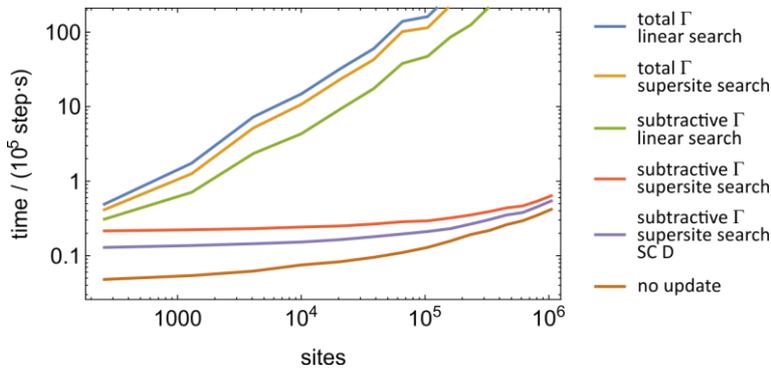

**Figure 6**: Total scaling for combinations of all presented algorithms from **Sections 3.1** to 3.3 on Fractal (ifort 11.1). Local update was employed in all tests.

However, if we replace both the accounting and search step by the Subtraction Scheme and the Supersite Search (red curve in **Figure 6**), we observe that both the scaling and overall performance improve dramatically. Because the Subtraction Scheme scales as $\mathcal{O}(1)$ and the Supersite Search scales as $\mathcal{O}(\sqrt{N})$, the overall scaling is now $\mathcal{O}(\sqrt{N})$. However, the $\sqrt{N}$ term is asymptotic for simulations with less than $10^5$ sites, resulting in practically constant run time per step even for medium to large simulation lattices. This is mostly due to the $\mathcal{O}(1)$ update step, which suffers from the lengthy literal evaluation of the CEH (Eq. 1). Replacing the energy function by the Supercluster Contraction (purple curve in **Figure 6**) decreases the time for small systems by about 40 % (cf. **Section 3.2**). This demonstrates that any single one of the proposed algorithms alone does not provide a great enhancement. To reach their full potential in VSSM, these three algorithms (Supercluster, Subtraction, Supersite) need to be combined (SSS-VSSM).

Finally, we examine how simulations with lateral interactions as implemented here perform compared to simulations without lateral interactions. To estimate the effect of removing lateral interactions from the KMC simulation we can simply deactivate the update step (by using the `no_update` function instead of `local_update` in the toy program) and pretend that updating does not take any time at all in simulations without lateral interactions (biasing this comparison in favor of simulations without lateral interaction). The result without update is plotted in **Figure 6** (brown curve) and shows again a significant decrease in total run time if no update is performed. For the smallest lattice size ($16 \times 16$) the speedup compared to local update with SCD (purple curve) is about 2.7, while for larger lattice sizes the difference is diminished as the $\mathcal{O}(\sqrt{N})$ search starts to affect the total run time. This final comparison indicates that, while the inclusion of lateral interactions comes at an additional cost, it can be made small by choosing appropriate algorithms.

We conclude that the computational cost in our test increases by less than a factor of 3 when including lateral interactions. However, our test does not involve a real reaction mechanism, pattern matching, rate computation, and there is only one energy evaluation per step and updated site. Real simulations come with a larger $\mathcal{O}(1)$ term in the update step, so that on the one hand, a more efficient energy evaluation will be even more impactful in small-scale simulations. On the other hand, the $\mathcal{O}(\sqrt{N})$ term from the search step will be even less noticeable. We conclude from this test that even for medium to large lattice sizes the run time of the update step governs the overall computational time, so that implementation of more sophisticated search algorithms is necessary only for larger system sizes. Instead, optimizing the run time of the update step (SC/CE-Hamiltonian, pattern matching, efficient rate calculation) takes priority in simulations with lateral interactions, and the supercluster contraction and subtraction scheme for energy updating (cf. **Section S1.2**) are practically feasible methods to do that.

## 4. Summary and conclusion

While lateral interaction models for reactions at surfaces have steadily gained popularity and grown in terms of complexity, their use in chemical kinetics has been limited by them being more computationally demanding than simulations without lateral interactions [2, 33]. This is in part due to the more elaborate Cluster Expansion Hamiltonian (CEH), which is a sum of several terms, in part due to the more extensive rate updating required in interacting coadsorbate systems, and in part due to the impracticality of the popular rate-catalog-based algorithms as the number of different rate constants increases exponentially with the number of interaction parameters. For site-based algorithms, on the other hand, it is recognized that the run time per step scales with the number of sites on the simulation lattice, which is equally undesirable.

Considering the computations required in performing the update, accounting, and search step in the Variable Step Size Method (VSSM) algorithm, we propose three ways to reduce the computational effort in KMC simulations with lateral interactions, two of them being new algorithms. 1. Represent the CE in a contracted form, by adding terms together into superclusters (Supercluster Contraction). The superclusters, each representing larger pieces of the whole CE, are pre-calculated, which significantly reduces the computational time spent on evaluating the lattice Hamiltonian during the simulation. 2. Computing only the change of the rate sum Γ, rather than recomputing it from scratch in every step, which reduces the overall scaling with respect to the number of sites. 3. Employing efficient search algorithms, such as the 2-Level [41] and K-Level [40] schemes (Supersite Algorithm).

In the standard site-based Variable Step Size Method (VSSM), the overall run time scaling with respect to the number of sites is determined by the accounting and search steps, both of which scale as $\mathcal{O}(N)$. The scaling of these steps is reduced to $\mathcal{O}(1)$ and $\mathcal{O}(N^{1/2})$ by the introduction of the Subtraction Scheme and Supersite Algorithm, respectively. The overall scaling with respect to the number of sites is now determined by the search step; however, even for medium to large lattice sizes ($<10^5$ sites), the update step, which includes the evaluation of the CEH, determines the overall run time. Using the Supercluster Approach reduces the overall time per step by 40 % or more. We have tested different Supercluster Hamiltonians for our CE models previously applied in our KMC simulations of the HCl and CO oxidation over $RuO_2$(110) [17, 18] and propose that the optimal supercluster contraction represents a compromise between the number of terms in the Hamiltonian and the size of the data set. However, the optimal solution appears to depend on the employed computer architecture, so that we cannot make general statements about the expected speedup and scaling of the Supercluster Approach at this point. Altogether, these algorithms constitute a significant improvement of KMC with lateral interactions compared to previous literature reports [2, 33].

The importance of lateral interactions in KMC models cannot be overstated, but the established algorithms currently favored in the catalysis field do not perform well with lateral interaction models. While complex lateral interaction models are still challenging in KMC simulations, the simulation algorithms presented here (Supercluster, Subtraction, Supersite) offer a way to overcome these obstacles. We have shown here that the computational cost increase by including lateral interactions in KMC can be small (around a factor of 3). This makes KMC simulations with lateral interactions in SSS-VSSM not only practically feasible, but very attractive for complex surface reactions.

**Supporting information**

Error handling in and additional applications of the Subtraction Scheme, optimal $s_x, s_y$ in the supersite algorithm. The sample programs (`supercluster.f90`, `subtraction.f90`, `supersite.f90`) are distributed via github (https://github.com/hessfran/SSS-VSSM/).

**Acknowledgment**

This project was funded by Deutsche Forschungsgemeinschaft through a research fellowship (HE-7782/1-1) and MIT Energy Initiative. Computational facilities were provided by the Yildiz group (Fractal) and NERSC (Cori). The author gratefully acknowledges Prof. Bilge Yildiz for guidance and mentorship, Prof. Herbert Over for advice and encouragement, and Dr. Lixin Sun and Dr. Pjotrs Zguns for helpful discussion and suggestions.